\documentclass[twocolumn]{aastex63}
\usepackage{graphicx}
\usepackage{setspace} 
\usepackage{amsmath} 
\usepackage[inline, shortlabels]{enumitem}

\def\be{\begin{equation}}
\def\bea{\begin{eqnarray}}
\def\eea{\end{eqnarray}}
\def\ee{\end{equation}}

\newcommand{\beq}{\begin{equation}}
\newcommand{\eeq}{\end{equation}}
\newcommand{\ba}{\begin{array}}
\newcommand{\ea}{\end{array}}

\submitjournal{ApJL}

\shorttitle{Radiation driven magnetic fields}
\shortauthors{Vyas and Pe'er}
\graphicspath{{./}{figures/}}

\begin{document}
\setstretch{1.0}
\title[Radiation-triggered magnetic fields around black holes]{Radiation-Driven Origin of Super-Equipartition Magnetic Fields in Accretion Discs and Outflows}

\correspondingauthor{Mukesh Kumar Vyas} 
\email{mukeshkvys@gmail.com}


\author{Mukesh Kumar Vyas}
\affiliation{Bar Ilan University, \\ Ramat Gan, Israel,
 5290002}
\author{Asaf Pe'er}
\affiliation{Bar Ilan University, \\ Ramat Gan, Israel,
 5290002}

\begin{abstract}
Magnetic fields play a central role in accretion physics around black holes, yet their physical origin within accretion flows remains an open problem.
In this work, we investigate the generation and subsequent evolution of magnetic fields triggered by anisotropic radiation fields in black hole accretion discs with compact rotating inner corona. We self-consistently evolve the magnetic field using the generalized field evolution MHD equation, including advection, shear-driven induction, and Hall effects. The radiation field acts as a primary field generator, while azimuthal rotation in the magnetized plasma provides rapid amplification.
We find that radiation-generated fields efficiently reach a dominant toroidal component by Keplerian rotation, leading to magnetic field strengths of order $\sim 10^{8}\,\mathrm{G}$ in the vicinity of a 10 solar mass black hole and accretion disc-corona emitting at luminosity equivalent to the Eddington unit. These magnetic fields are achieved within viscous timescales and reach or exceed local equipartition estimates based on gas pressure. When vertical outflows are included, the amplified magnetic fields are advected into the corona, magnetizing disc-launched winds and jet precursors with field strengths of similar order.
Our results demonstrate that radiation is not merely a passive component of accretion flows, but provides a robust and unavoidable trigger for the generation of dynamically significant magnetic fields. Our results provide a physically grounded explanation for the origin of large-scale, structured magnetic fields in and around accretion discs. This mechanism offers a pathway for magnetizing accretion discs and their outflows without invoking externally supplied magnetic flux, with broad implications for X-ray binaries, active galactic nuclei and other transients such as gamma-ray bursts (GRBs).
\end{abstract}
\keywords{High energy astrophysics; Theoretical models, Plasma astrophysics, Accretion discs}

\section{Introduction}

Magnetic fields are a central ingredient of accretion physics around compact objects, governing angular momentum transport, disc turbulence, jet launching, and the coupling between radiation and plasma \citep{1988ASSL..133.....R,2002RvMP...74..775W,2002PhT....55l..40K,2005PPCF...47A.205S}. Observations and numerical simulations of black hole accretion systems consistently indicate the presence of strong, large-scale magnetic fields in the innermost disc regions, where relativistic jets and powerful winds are launched (e.g., \citealt{1991ApJ...376..214B,1995ApJ...440..742H,2020ARA&A..58..407D,2021NewAR..9201610K,2022MNRAS.511.3795N,2023ApJS..264...32B}). Despite their importance, the physical origin of such magnetic fields in accretion discs remains an open and longstanding problem.
As of today, 
most contemporary models rely on the amplification of weak seed fields by magnetorotational instability (MRI) turbulence or large-scale dynamo action \citep{1978mfge.book.....M,1991ApJ...376..214B,1995ApJ...440..742H}. 

A {different,} physically well-motivated class of such mechanisms arises from the non-conservative nature of radiation forces in astrophysical plasmas, namely situations where radiation force has nonvanishing curl \citep[$\nabla \times \mathbf{F}_{\rm rad} \neq 0$  e.g.,][]{2010ApJ...716.1566A,2014ApJ...782..108S}. In this case, radiation induces charge separation, drives electric currents, and generates magnetic fields. One early realization of this idea is the Poynting--Robertson effect, in which radiation drag on orbiting electrons in an accretion disc produces toroidal currents \citep{1977A&A....59..111B}. Subsequent generalizations incorporating finite conductivity and extended illumination geometries led to the development of the cosmic battery framework \citep{1998ApJ...508..859C,2002ApJ...580..380B,2006ApJ...652.1451C,2008ApJ...674..388C,2015ApJ...805..105C}.

These studies demonstrated that radiation-driven batteries can operate robustly over long timescales and in diverse astrophysical environments, including accretion discs {and seed magnetic field generation around stars in the early Universe} \citep{2003PhRvD..67d3505L,2010ApJ...716.1566A}. However, for accreting black holes, the generated magnetic fields were found to remain weak, typically $\lesssim 10^{2}$~G over viscous timescales, requiring unrealistically long growth times to approach equipartition with the gas or radiation energy density \citep{1977A&A....59..111B,2002ApJ...580..380B}. {Even with improved modelling, this cosmic battery mechanism for X-ray discs yields magnetic field strengths several orders of magnitude below equipartition within realistic accretion lifetimes }
\citep{2014ApJ...794...27K,2015ApJ...805..105C}.

Most existing radiation-battery calculations assume a luminous Keplerian disc extending beyond the innermost stable circular orbit (ISCO), while neglecting the possible presence of a compact inner corona. The existence of such a corona has been a requirement to explain X-ray observations from X-ray binaries \citep{1976SvAL....2..191B,1991ApJ...380L..51H}.
{Further, the observations indicate that transient jet ejections are triggered when the system enters the very high/intermediate state (VHS/IS) at X-ray luminosities 
$L_{X,\mathrm{VHS}} \sim 0.1$--$1\,L_{\rm Edd}$ 
Sources such as GX 339$-$4, XTE J1550$-$564, XTE J1859$+$226, and GRO J1655$-$40 cluster in this luminosity range, while GRS 1915$+$105 reaches luminosities close to $L_{\rm Edd}$. 
This supports a picture in which jet launching is associated with a luminous, geometrically compact (extending a few gravitational radii) inner corona \citep{2004MNRAS.355.1105F,Miniutti2004, 2012A&A...537A..18M}.
}

In our recent work \citep{2025ApJ...988L..59V}, we revisited radiation-driven magnetic field generation by explicitly computing the full three-dimensional radiation field produced by a Keplerian disc and a compact inner corona. We demonstrated that the presence of a luminous, rotating corona dramatically enhances $\nabla \times \mathbf{F}_{\rm rad}$, leading to magnetic field generation reaching $\sim 10^{5}$~G within the accretion timescales in stellar-mass black hole systems, several orders of magnitude larger than classical Poynting--Robertson estimates. These fields were found to be one order of magnitude below the equipartition level.
\cite{2025ApJ...988L..59V} focused exclusively on the radiation source term in the induction equation and did not follow the subsequent dynamical evolution of the generated fields. {The study was relevant only for static and high-density plasma above the disc plane, allowing us to ignore induction and Hall effects.}

In this work, we take the next critical step by self-consistently evolving the magnetic field using the full induction equation, including advection, shear-driven induction, and Hall effects, starting from radiation-generated magnetic fields. By constructing the full three-dimensional magnetic-field structure resulting from radiation-driven generation and subsequent MHD evolution, we provide a coherent picture of disc and coronal magnetization.
%

Our results show that radiation-generated fields, when coupled to differential rotation, are rapidly amplified to strengths of $\sim 10^{8}\,\mathrm{G}$ within viscous timescales for Eddington-limited accretion in stellar-mass black holes, exceeding equipartition estimates based solely on gas pressure. In this sense, radiation acts as a robust and unavoidable trigger for the generation of dynamically dominant magnetic fields in accretion discs, while shear provides the primary channel for amplification.

In addition, we find that radiation-generated magnetic fields are efficiently advected into vertically outflowing plasma, leading to substantial magnetization of disc-driven winds and jet precursors. In this regime, magnetic field strengths of order $\sim 10^{7}\,\mathrm{G}$ are sustained on scales of several tens of gravitational radii above the disc plane, demonstrating that accretion-disc radiation can magnetize not only the disc itself but also the emerging outflows.

The emergence of strong toroidal fields through shear-driven induction, accompanied by large-scale poloidal components penetrating the disc, provides a natural pathway for jet and wind launching without requiring externally supplied magnetic flux. \textit{Our results, therefore, offer a physically grounded answer to a long-standing question: where do the large-scale, structured magnetic fields in and around accretion discs originate?} The implications are broad, spanning X-ray binaries, active galactic nuclei, and radiation-dominated accretion flows.

This paper is organized as follows. In Section \ref{sec_mat_num} we present the governing equations for magnetic field evolution, including radiation, induction, and Hall effects, along with the numerical setup and radiation field prescription. Section \ref{sec_res} presents the results of the fully coupled evolution and discusses the emergence of strong disc-penetrating magnetic fields. We conclude in Section \ref{sec_conc} with implications and future directions.

\section{Mathematical Framework and Computational Scheme}
\label{sec_mat_num}

\begin {figure}
\begin{center}
 \includegraphics[width=9cm, angle=0]{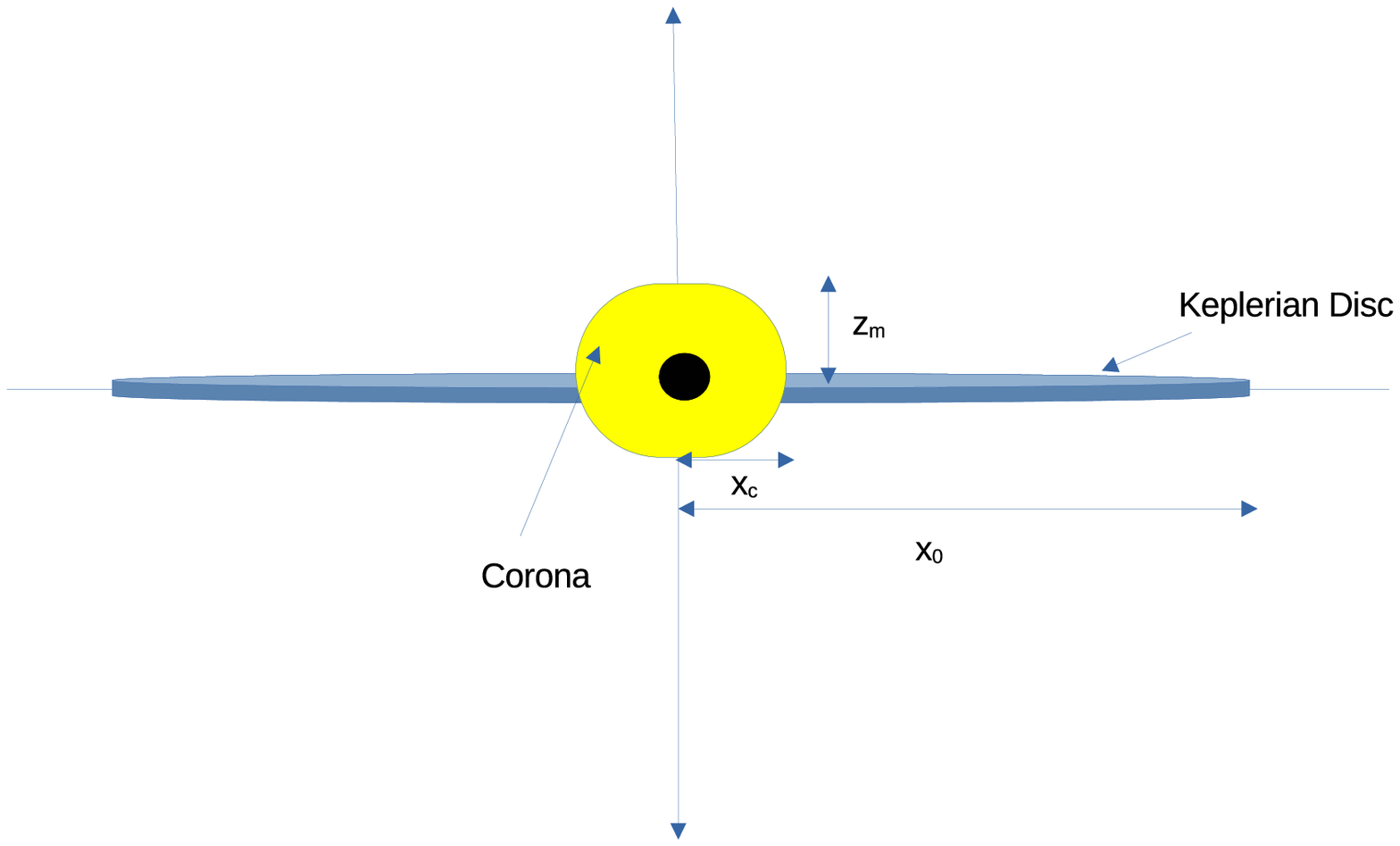}
\caption{Schematic illustration of the  adopted disc--corona geometry.
A geometrically thin Keplerian disc extends radially from $x_c$ to $x_0$ situated at $z = 0$,
while a compact, opaque corona occupies an oblique region of radius $x_c$ and vertical
extent $z_m$ above the black hole.
The corona emits isotropically and is treated as optically thick, so radiation entering
its interior is absorbed.
This geometry is used to compute the anisotropic radiation field and the resulting
radiation--driven magnetic source term as shown in the appendix.}
\label{lab_accretion_geom}
 \end{center}
\end{figure}

We study the radiation–driven evolution of magnetic fields in a plasma surrounding an accreting compact object, adopting a non-relativistic MHD framework in cylindrical coordinates $(r,\phi,z)$. 
Special relativistic effects are included only through Doppler boosting of radiation emitted by rotating surfaces, while relativistic corrections to the plasma equations are ignored. The radiation field is computed independently from the magnetic field evolution and treated as an externally imposed driver; feedback of the generated magnetic field on the radiation field is neglected. Emission is assumed to originate exclusively from geometrically thin surfaces: the disc photosphere and a single inclined coronal emitting surface that connects the disc--corona interface to the upper coronal region (Figure \ref{lab_accretion_geom}). The disc body and the interior of the corona are treated as effectively opaque, such that photons are absorbed upon entering these regions, and no volume emissivity or radiative transfer through the coronal interior is included.

Under these assumptions, the evolution of the magnetic field is governed by the 
Hall–MHD-induction equation with a radiation source term as \citep{2010ApJ...716.1566A,2014ApJ...782..108S, 2025ApJ...988L..59V},
\begin{equation}
\frac{\partial \mathbf{B}}{\partial t}
=
\nabla \times (\mathbf{v} \times \mathbf{B})
-
\nabla \times
\left[
\frac{c}{4\pi e n_e}
\left[ (\nabla \times \mathbf{B}) \times \mathbf{B} \right]
\right]
+
\mathbf{F}_0(r,z),
\label{eq:hall_induction}
\end{equation}
where $\mathbf{B}$ is the magnetic field, $\mathbf{v}$ is a prescribed background velocity field, $n_e$ is the electron number density, $e$ is the elementary electron's charge and $c$ is light speed. The electron resides in an overall force-free field. The radiation–driven source term $\mathbf{F}_0(r,z)$ arises from the curl of the radiation force acting on electrons and is given by
\begin{equation}
\mathbf{F}_0(r,z)
\equiv
-\frac{c}{e}\,\nabla \times \mathbf{f}_{\rm rad}(r,z),
\end{equation}
where $\mathbf{f}_{\rm rad}$ is the radiation force per electron. In the Thomson scattering regime relevant here, the radiation force is proportional to the local radiation flux,
\begin{equation}
\mathbf{f}_{\rm rad} = \frac{\sigma_{\rm T}}{c}\,\mathbf{F}_{\rm rad},
\end{equation}
$\sigma_{\rm T}$ is Thomson scattering cross section. The source term can be written explicitly in terms of spatial derivatives of the radiation flux,
\begin{equation}
\mathbf{F}_0(r,z)
=
-\frac{\sigma_{\rm T}}{e}\,
\nabla \times \mathbf{F}_{\rm rad}(r,z).
\label{eq:F0_def}
\end{equation}
Thus, magnetic fields are generated wherever the radiation flux possesses a non–zero curl, reflecting the intrinsic anisotropy of the radiation field produced by the disc–corona geometry.
The detailed construction of the radiation flux components and their spatial derivatives leading to estimation of $\nabla \times \mathbf{F}_{\rm rad}(r,z)$ is described in Appendix~\ref{sec_app}. Throughout this work, we assume a non-rotating (Schwarzschild) black hole, so that the Keplerian disc is truncated at $x_{\rm in} \simeq x_c \ge  3 r_g$. 
Resistive diffusion is neglected in the present study, corresponding to the ideal limit (zero resistivity - $\eta = 0$).

The three terms on the right–hand side of Eq.~(\ref{eq:hall_induction}) represent, respectively:
(i) magnetic induction and advection by the flow, (ii) the Hall current contribution, which introduces dispersive and nonlinear magnetic coupling, and
(iii) radiation–driven magnetic–field generation arising from the non–conservative nature of the radiation force.
The solenoidal constraint,
\begin{equation}
\nabla \cdot \mathbf{B} = 0,
\end{equation}
is preserved numerically throughout the evolution and monitored as a diagnostic.




\subsection{Considered geometry of disc-corona}
The radiation field is computed by numerical integration of disc and corona specific intensities (as shown in Appendix \ref{sec_app}) over the disc and corona surfaces for a given field point above the disc plane (Figure \ref{lab_accretion_geom}). These fluxes are stored in a table, which is then interpolated over the numerical domain to solve Equation \ref{eq:hall_induction}. The solution of this equation gives the magnetic field generation and evolution of plasma in the vicinity, whose properties are discussed in section \ref{sec_plasma} below.

The Keplerian disc is assumed to accrete at $0.1$ times the Eddington accretion rate (corresponding to sub-Eddington luminosities), while the corona radiates with a total luminosity of one Eddington. Hence, the radiation from the corona dominates over the Keplerian disc.
The radiation field is mapped onto a two–dimensional cylindrical domain $(r,z)\in[1.1,10.1]\,r_g\times[0.1,10.1]\,r_g$. A $10\,M_\odot$ black hole is assumed in the conversion to physical units. The corona occupies a compact region of characteristic radial size $x_c$ and vertical extent $z_m$, while the Keplerian disc extends to larger radii $x_0 = 40 r_g$. The disc rotates with Keplerian angular velocity, whereas the corona is assumed to undergo solid–body rotation, with the rotational speed at the disc–corona interface matched to the Keplerian speed at $r=x_c$. 

\subsection{Considered plasma model}
\label{sec_plasma}
The formulation allows us to track the three-dimensional magnetic-field evolution resulting from radiation-magnetization in accretion environments, while explicitly capturing: 
(i) amplification by rotational shear, 
(ii) redistribution of magnetic fields into initially field-free regions such as the disc outer surface, and 
(iii) nonlinear Hall--induced structuring and dispersive dynamics.

The radiation from the disc corona components shown above generates time-dependent magnetic fields in the tenuous plasma in the vicinity. The plasma is modelled as a dilute, quasi-neutral medium with a prescribed and spatially uniform electron number density, $n_e$, which is held constant throughout the simulations. 
For the optically thin region surrounding the compact corona, we adopt a Thomson optical depth $\tau \simeq 10^{-4}$. 
Using $\tau = n_e \sigma_T H$ with $H \sim \simeq 10^6\,\mathrm{cm}$ for a $10\,M_\odot$ black hole gives 
$n_e \simeq \tau / (\sigma_T H) \approx 10^{14}\,\mathrm{cm^{-3}}$, 
which we take as a representative density for the tenuous plasma outside the corona.
The velocity field includes a weak vertical outflow with speed $v_z=0$ for magnetization of the immediate surroundings of the disc-corona and $v_z=f(z)$ (Equation \ref{eq_vz}) for magnetization of outflows. The azimuthal component is inherited from the Keplerian rotation,

\begin{equation}
\mathbf{v}(r,\phi,z,t) = 
\left(
0,\;
v_\phi(r),~v_z(z,t)
\right),
\end{equation}

The plasma velocity field is prescribed and not evolved self-consistently, allowing controlled exploration of magnetic transport and amplification processes. 
As an approximation, the radial inflow component is neglected as it is much smaller compared to other components

We consider two distinct cases within this unified framework:

\begin{enumerate}

\item \textit{Disc magnetization without vertical outflow.}  

In this configuration, the vertical velocity component is set to zero everywhere above the disc, $v_z = 0$. Plasma motion is therefore confined to azimuthal rotation and slow radial inflow, with the latter governed by viscous angular momentum transport within the disc. The characteristic residence time of matter at cylindrical radius $x$ is set by the viscous timescale,
\begin{equation}
t_{\rm visc} \equiv \frac{x}{v_r},
\end{equation}
where $v_r$ is the radial inflow velocity. Following the standard viscous-disc prescription \citep{1977A&A....59..111B}, the radial velocity can be written as,
\begin{equation}
v_r \simeq 7.7 \times 10^{9}\,
\alpha\, \dot{m}^{-1}\,
\frac{1 - x'^{-1/2}}{x'^{5/2}}
\ \mathrm{cm\ s^{-1}},
\end{equation}
where $\alpha$ is the viscosity parameter, $\dot{m}$ is the accretion rate in Eddington units, and $x' = x/r_g$ is the cylindrical radius of the disc in Schwarzschild units. 

For values considered in this paper, $\alpha \sim 0.1$, $\dot{m} = 0.1$, and representative distances from black hole $x \sim 3$--$5 \, r_g$, this yields a viscous time-scale
\begin{equation}
t_{\rm visc} \sim \mathcal{O}(1)\ {\rm s},
\end{equation}

\item \textit{Magnetization of outflowing plasma.}  
In this case, a vertically accelerating outflow is included. The vertical velocity is prescribed as a smooth, height-dependent function,
\begin{equation}
v_z(z) = v_{\rm i} + v_{z,\max}\,\tanh\!\left(\frac{z-z_l}{H}\right),
\label{eq_vz}
\end{equation}
where $z_l$ denotes the disc photosphere height, $H$ is a characteristic acceleration scale, and $v_{\rm i} \ll c$ ensures a finite velocity at the disc surface. This profile yields slow acceleration close to the disc, followed by gradual saturation at sub-relativistic speeds, allowing plasma to remain in the magnetization region for a finite time before being advected vertically.

The adopted vertical velocity profile is not intended as a self-consistent wind solution, but rather as a phenomenological representation of the slow-launch and extended acceleration behaviour commonly found in thermally driven \citep{1958ApJ...128..664P}, and radiation-driven disc outflows \citep{2019MNRAS.482.4203V, 2021MNRAS.501.4850R}. This prescription allows controlled exploration of how radiation-generated magnetic fields are advected, redistributed, and sustained on larger scales above the disc plane.

In this regime, the relevant timescale for the magnetization is set by the vertical advection timescale,
\begin{equation}
t_{\rm adv} \simeq \int_{0}^{H} \frac{dz}{v_z(z)},
\end{equation}
which directly controls the duration over which radiation-generated magnetic fields can grow before being transported to larger heights. Compared to the viscous case, this finite advection time limits the maximum attainable field strength near the disc surface, while simultaneously enabling the transport of strong magnetic fields into the corona and outflow region. For typical parameters considered in this work, $v_{\rm i}=10^{-10}c$, $H=25 r_g$ and $v_{z,\max}=0.05c$, we obtain $t_{\rm adv}\sim \mathcal{O}(1)\ $s. The choices of these parameters imply typical length scales above the disc plane and disc winds with moderate speeds.
\end{enumerate}

These obtained values of $t_{\rm visc}$ and $t_{\rm adv}$ define the limit of time integration of equation \ref{eq:hall_induction}.
Apart from the presence or absence of vertical advection, all other physical assumptions, boundary conditions, and numerical treatments are identical between the two cases. This controlled setup enables a direct comparison between disc-dominated and outflow-dominated magnetization regimes, and isolates the role of magnetization timescale in determining the strength, spatial extent, and dynamical impact of radiation-generated magnetic fields.

\subsection{Numerical Method}

Equation~(\ref{eq:hall_induction}) is solved in cylindrical coordinates using Mathematica’s \texttt{NDSolve} framework with the Method of Lines \citep{schiesser2012numerical}.
The magnetic field is decomposed into its cylindrical components $(B_r, B_\phi, B_z)$, yielding three coupled, nonlinear partial differential equations that are evolved self–consistently.

Spatial discretization is performed on a tensor–product grid in $(r,\phi,z)$, with uniform resolution in the radial and vertical directions and periodic boundary conditions in the azimuthal direction.
Time integration is fully adaptive, with the instantaneous timestep recorded throughout the evolution to monitor numerical stability and stiffness mainly introduced by Hall term.

The system is initialized with a vanishing magnetic field,
\begin{equation}
\mathbf{B}(t=0) = 0,
\end{equation}
so that all subsequent magnetic structure arises self–consistently from radiation forcing, induction, and Hall dynamics.

Zero–Dirichlet boundary conditions are imposed on the magnetic field at the radial and vertical domain boundaries, while periodicity is enforced in $\phi$.
Diagnostics are performed to verify that numerical divergence errors remain small compared to $|\mathbf{B}|/L$ at all times, where $L$ is the local grid scale.




\begin{figure*}
\begin{center}
\includegraphics[width=18cm, angle=0]{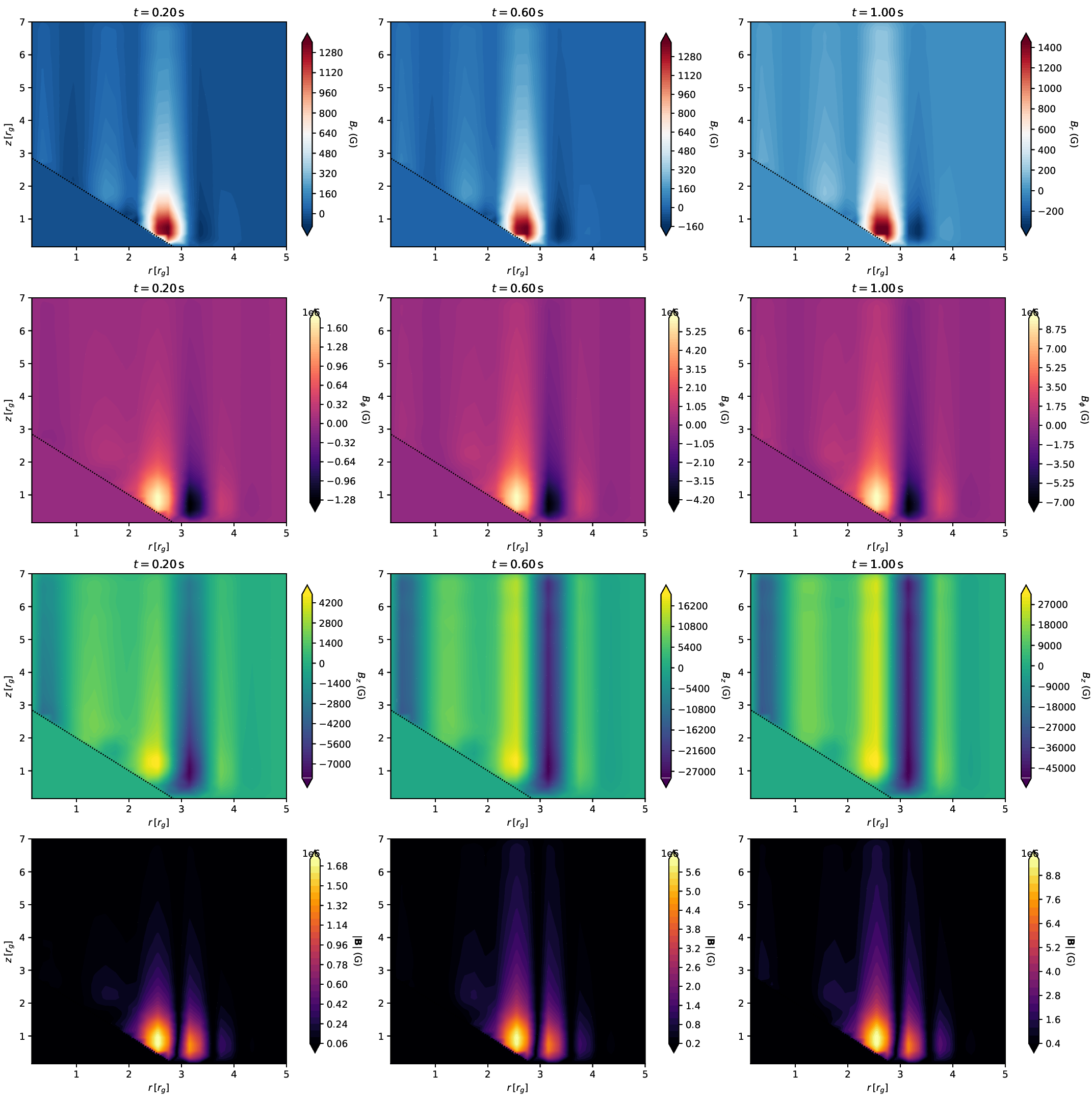}
\caption{Time evolution of the magnetic field components for a vertically outflowing plasma following the velocity profile given by equation~\ref{eq_vz}. The three columns correspond to times $t=0.2$, $0.6$, and $1.0\,\mathrm{s}$ (advective timescale). The top three rows show the radial ($B_r$), azimuthal ($B_\phi$), and vertical ($B_z$) components, while the bottom row shows the magnetic field magnitude $|\mathbf{B}|$. The radiation source term $\mathbf{F}_0(r,z)$ is obtained from precomputed radiative transfer tables for a luminous disc--corona system around a $10\,M_\odot$ black hole. Magnetic fields are initialized to zero and evolve under the combined action of radiation forcing, shear-driven induction, and the Hall term. The polarity reversal of magnetic fields above $r=3r_g$, especially in the azimuthal component, is caused by the interplay between the radiation term and the induction term. The black dotted line marks the geometrical extension of the corona.
}
\label{lab_Gen_res}
 \end{center}
\end{figure*}

\begin{figure*}
\begin{center}
\includegraphics[width=18cm, angle=0]{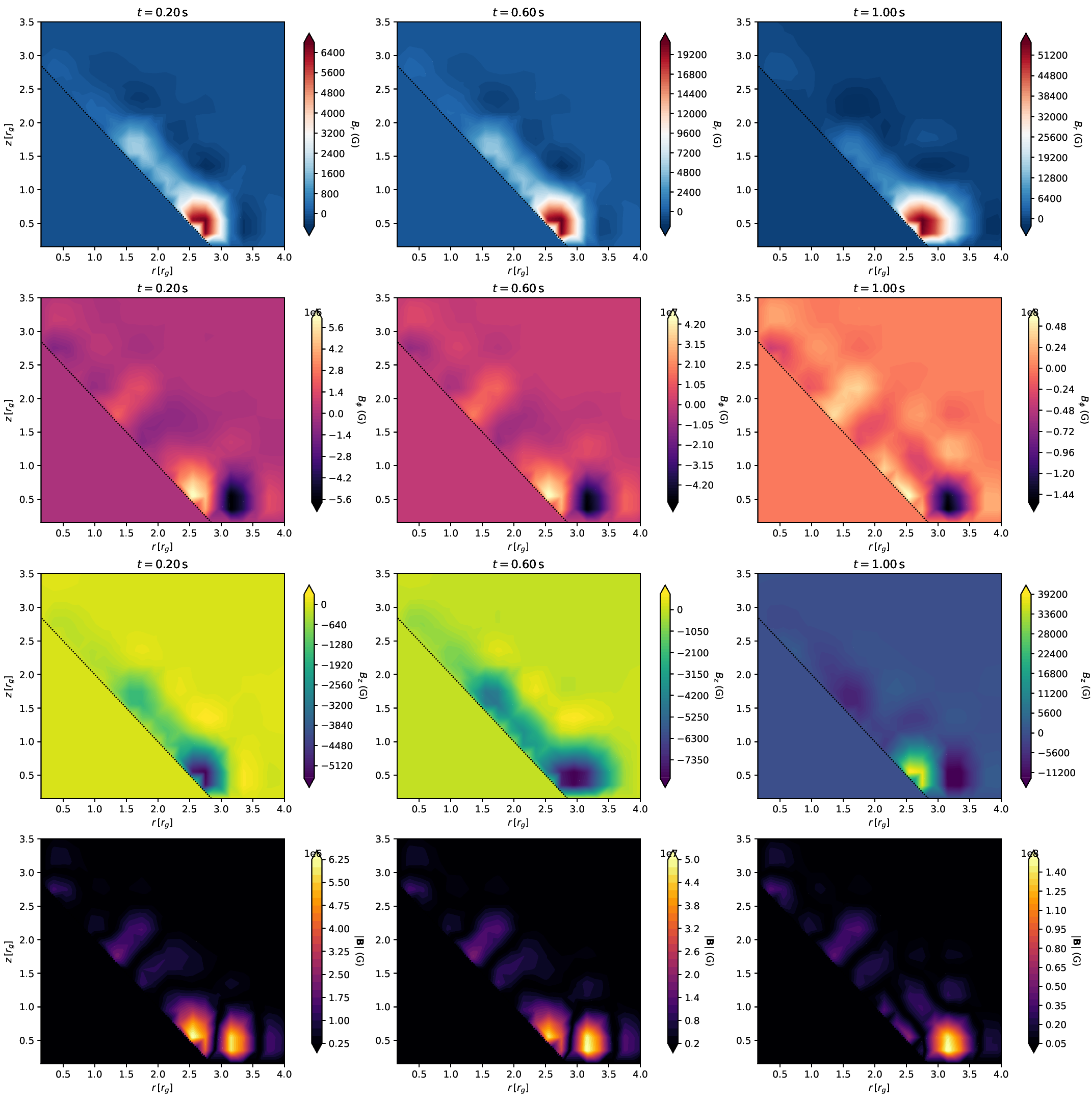}
\caption{Same as Figure \ref{lab_Gen_res} but with $v_z=0$, which implies the magnetization of the accretion disc upper surface due to disc-corona radiation field. In absence of vertical velocity component, the magnetization in plasma remains confined to the region close to the disc plane. The maximum runtime $1$ second corresponds to the typical viscous timescale.}

\label{lab_Gen_res_disk}
 \end{center}
\end{figure*}

\begin{figure}
\begin{center}
 \includegraphics[width=8.5cm, angle=0]{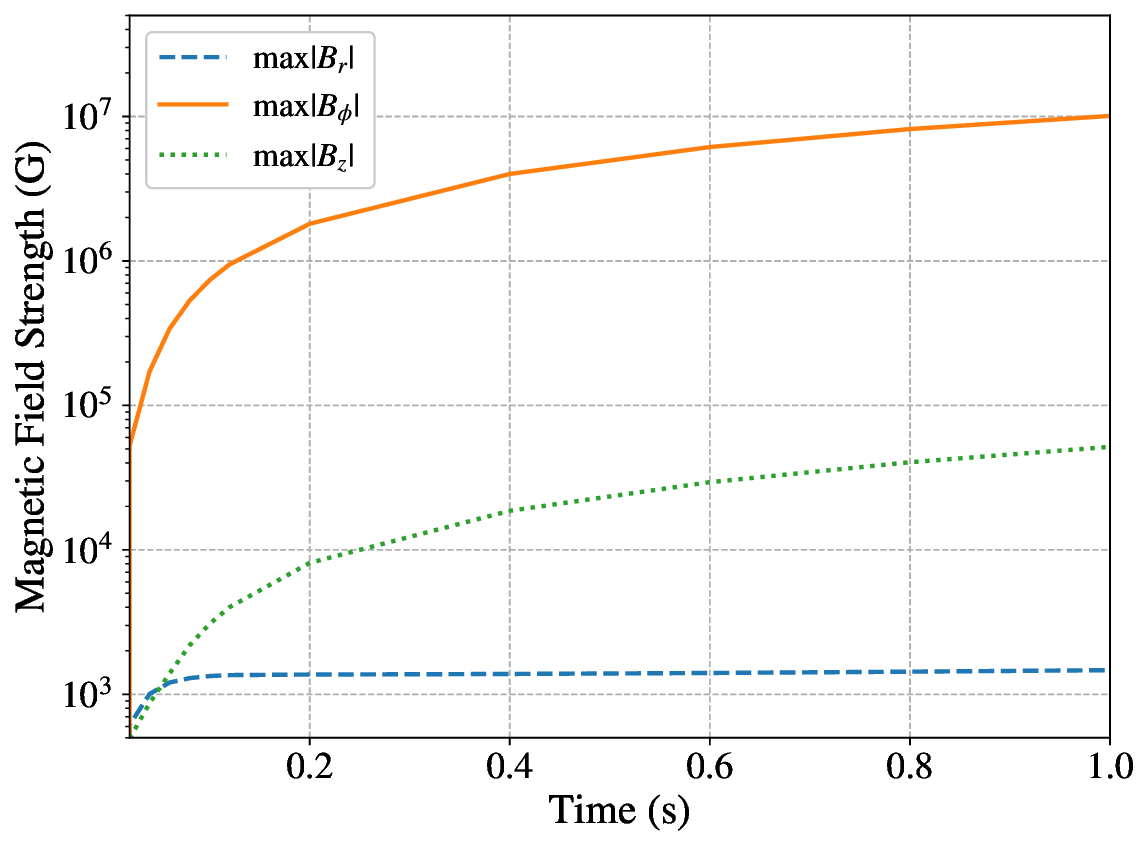}
\includegraphics[width=8.5cm, angle=0]{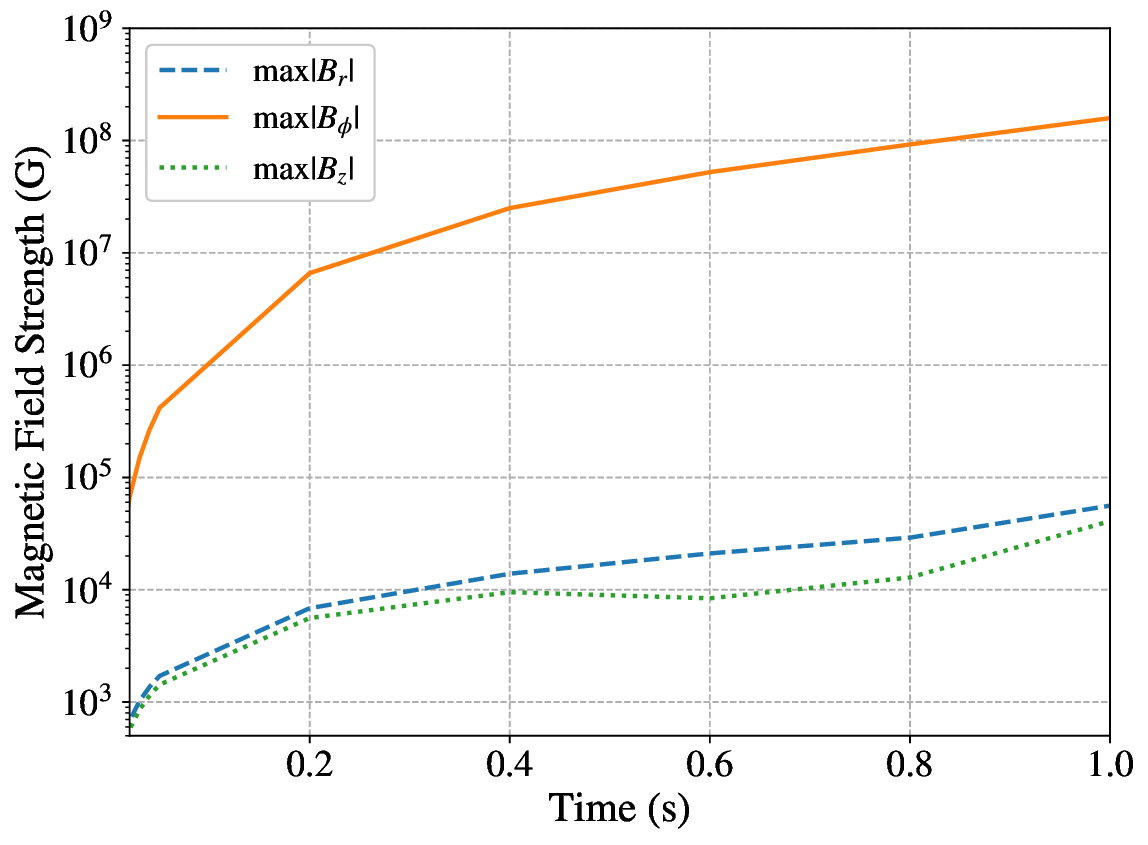}
\caption{The evolution of maximum magnetic field components for magnetized outflows in Figure \ref{lab_Gen_res} (top panel) and magnetized disc surface in Figure \ref{lab_Gen_res_disk} (bottom panel)}
\label{lab_max_mag_from_combined_files}
 \end{center}
\end{figure}

\section{Results and Interpretation}
\label{sec_res}

Figures~\ref{lab_Gen_res} and \ref{lab_Gen_res_disk} present the time evolution of the magnetic field components in the $(r,z)$ plane for the two physical regimes considered in this work: magnetization of vertically outflowing plasma (Figure \ref{lab_Gen_res}) and magnetization of the disc upper surface in the absence of vertical motion (Figure \ref{lab_Gen_res_disk}). For the coronal extension, we consider $x_c=z_m=3 r_g$ and the Keplerian disc lies at plane $z_l=0$ (ideally thin disc).
In both figures, the three columns correspond to representative times $t = 0.2$, $0.8$, and $1.0\,\mathrm{s}$. For a ten solar mass black hole considered here, this timescale is around $2\times 10^4$ gravitational timescale ($2GM_{BH}/c^3$), which shows that the viscous timescales for the accreted matter and advective timescales for slowly evolving outflows are much longer. The top three rows show the radial ($B_r$), azimuthal ($B_\phi$), and vertical ($B_z$) components of the magnetic field, while the bottom row displays the magnitude of the magnetic field, $|\mathbf{B}| = \sqrt{B_r^2 + B_\phi^2 + B_z^2}$.

Figure~\ref{lab_Gen_res} illustrates the case with a vertical outflow, in which the plasma above the disc is advected upward with a height dependent velocity profile. Magnetic field generation is initially localized to the region immediately outside the optically thick disc and corona, where the radiation--driven source term $\mathbf{F}_0(r,z)$ is non-zero. Weak poloidal magnetic field is produced near the disc--corona interface and along the inclined coronal surface, reflecting the anisotropic radiation geometry. Once azimuthal shear is present, these initial fields rapidly develop a dominant toroidal component. As time progresses, the magnetic field is advected vertically with the outflow, leading to sustained magnetization of the plasma on scales extending several gravitational radii above the disc plane. The resulting field configuration is spatially asymmetric and concentrated along the coronal boundary, highlighting the role of the coronal geometry in shaping the magnetic structure.

Figure~\ref{lab_Gen_res_disk} shows the complementary case in which the vertical velocity is set to zero, $v_z = 0$, corresponding to magnetization of the plasma without mass outflow. In this regime, the radiation-generated magnetic fields remain confined to the vicinity of the disc--corona surface. The contribution from the Keplerian disc in producing the field is negligible compared to the luminous corona. Thus, the coronal contribution dominates, and the fields remain confined in the region close to the coronal region. 
The absence of vertical advection prevents upward transport of magnetic flux, allowing the field to accumulate and intensify locally. As in the outflowing case, azimuthal shear efficiently amplifies the radiation-generated poloidal field into a strong toroidal component. 
Notably, both Figures~\ref{lab_Gen_res} and \ref{lab_Gen_res_disk} exhibit a clear polarity reversal in the toroidal magnetic field component, $B_\phi$. The physical origin of the strong $B_\phi$ amplification and the emergence of polarity reversals is explored in detail in Section~\ref{ref_sec_interpret} below.

\begin{figure}
\begin{center}
 \includegraphics[width=8.5cm, angle=0]{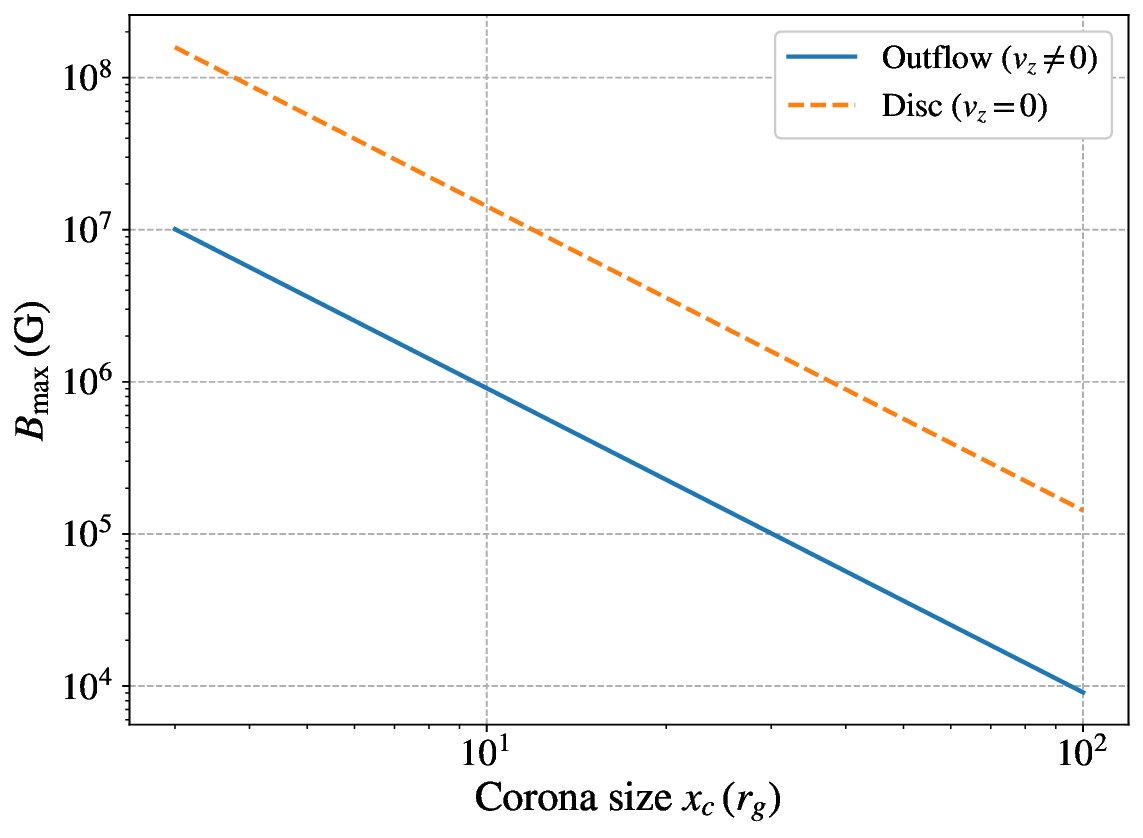}
\includegraphics[width=8.5cm, angle=0]{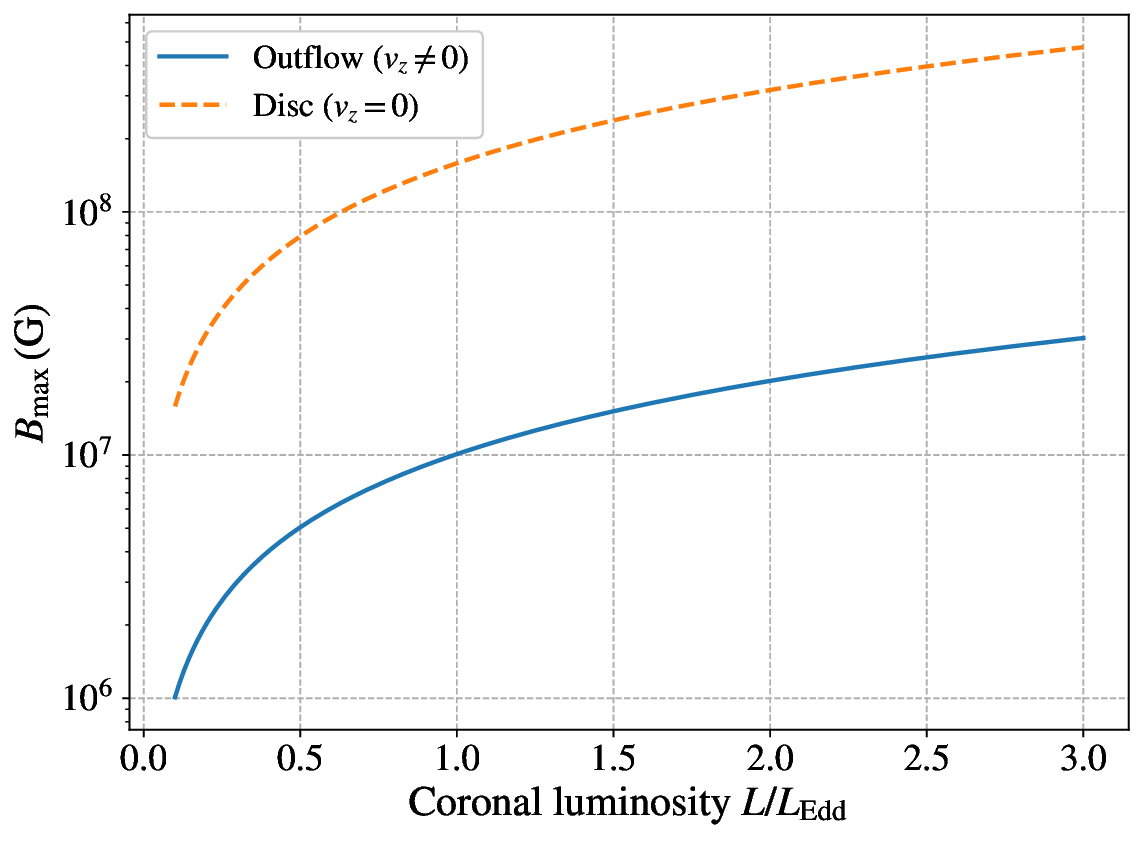}
\caption{Variation of Maximum magnitudes of magnetic field as shown in Figure \ref{lab_max_mag_from_combined_files} at $t=0$ for disc outflows (blue solid) and disc surface (dashed magenta) with corona size (top panel) and corona luminosity (bottom panel). The obtained relations are $B_{max}\propto 1/x_c^2$ and $B_{max}\propto L_c$
}
\label{lab_lab_max_mag_with_xc_and_l}
 \end{center}
\end{figure}

The temporal evolution of the maximum magnetic field components for both regimes is summarized in Figure~\ref{lab_max_mag_from_combined_files}. The top panel shows the evolution for the outflowing case, while the bottom panel corresponds to the disc--magnetization case. In both scenarios, the growth of the total magnetic field is dominated by the toroidal component $B_\phi$, which increases by several orders of magnitude within $\sim 1\,\mathrm{s}$. In the outflowing case, the peak magnetic field reaches $\sim 10^{7}\,\mathrm{G}$, demonstrating efficient magnetization of disc--launched outflows. In contrast, when vertical advection is suppressed, the magnetic field at the disc surface grows more rapidly and attains strengths of order $\sim 10^{8}\,\mathrm{G}$, reflecting prolonged shear amplification in a confined region.
The obtained magnetic fields strongly depend upon the size of the corona and its overall luminosity. In the previous case, the optimum values are chosen. The variation of produced $B_{max}$ with corona size $x_c$ and luminosity ($L_c$) is shown in Figure \ref{lab_lab_max_mag_with_xc_and_l} at $t=1$s, and as expected, the fields drop as inverse square with corona size and linearly increase with the luminosity (see appendix \ref{sec_app} for explanation ).

\subsection{Interpretation of high magnitudes and alternating polarity of $B_\phi$.}
\label{ref_sec_interpret}

When an azimuthal flow component is included, the magnetic-field evolution changes
qualitatively even in the absence of the Hall term.
To isolate this effect, we
consider the induction equation, including both the radiation source term and the
ideal induction term,
\begin{equation}
\frac{\partial \mathbf{B}}{\partial t}
=
\nabla \times (\mathbf{v} \times \mathbf{B})
+
\mathbf{F}_0,
\label{eq:induction_radiation}
\end{equation}
and consider an axisymmetric flow with purely azimuthal velocity,
\(
\mathbf{v} = (0, v_\phi(r), 0).
\)
This reduction applies to both physical regimes
considered in this work, namely magnetization of the disc surface in the absence of
vertical motion and magnetization of vertically outflowing plasma, since the
azimuthal shear is common to both cases.

Evaluating the curl explicitly in cylindrical coordinates and noting that
$\partial/\partial\phi = 0$, one finds that the radial and vertical components
satisfy
\begin{equation}
\frac{\partial B_r}{\partial t} = F_{0,r},
\qquad
\frac{\partial B_z}{\partial t} = F_{0,z},
\end{equation}
while the azimuthal component evolves according to
\begin{equation}
\frac{\partial B_\phi}{\partial t}
=
(\nabla \times (\mathbf{v} \times \mathbf{B}))_\phi
+
F_{0,\phi}.
\end{equation}
For a purely azimuthal flow and the axisymmetric limit ($\partial/\partial\phi = 0$) of the induction equation, the induction term reduces to
\begin{equation}
(\nabla \times (\mathbf{v} \times \mathbf{B}))_\phi
=
v_\phi
\left(
\frac{\partial B_r}{\partial r}
+
\frac{\partial B_z}{\partial z}
\right)
+
B_r \frac{d v_\phi}{dr}.
\end{equation}
Using the solenoidal constraint $\nabla\cdot\mathbf{B}=0$ under axisymmetry,
\begin{equation}
\frac{1}{r}\frac{\partial}{\partial r}(rB_r)
+
\frac{\partial B_z}{\partial z}
=0,
\end{equation}
the derivative terms cancel identically, and the evolution equation simplifies to
\begin{equation}
\frac{\partial B_\phi}{\partial t}
=
-\,\frac{v_\phi}{r}\,B_r
+
B_r\frac{d v_\phi}{dr}
+
F_{0,\phi}.
\label{eq:Bphi_full}
\end{equation}
For a Keplerian azimuthal velocity $v_\phi\propto r^{-1/2}$, this further reduces to
\begin{equation}
\frac{\partial B_\phi}{\partial t}
=
-\frac{3}{2}\frac{v_\phi}{r}\,B_r
+
F_{0,\phi}.
\label{eq:Bphi_shear}
\end{equation}

Equation~(\ref{eq:Bphi_shear}) reveals two physically distinct roles played by the
radiation field and the azimuthal velocity component. 
The radiation source term $F_{0,\phi}$
provides the generation and local sign of the toroidal magnetic field.

The Keplerian shear acts on the
radiation--generated radial field $B_r$ to produce rapid toroidal amplification.
This shear term dominates the long--time evolution and is responsible for the
orders of magnitude increase in $|B_\phi|$ observed in both the disc--magnetization
and outflow--magnetization cases. Importantly, because the shear term is
proportional to $B_r$, any spatial sign variation in the radiation-generated radial
field is directly mapped into alternating polarity of $B_\phi$. The resulting
polarity reversal, visible in both Figures~\ref{lab_Gen_res} and
\ref{lab_Gen_res_disk}, is therefore not generated by radiation alone, but emerges
from the coupling between the radiation-seeded poloidal field and azimuthal
rotation. The presence or absence of vertical advection does not alter this
mechanism, but instead controls the subsequent spatial redistribution of the
amplified toroidal field.

The rapid amplification of the toroidal magnetic field can be understood with a
simple order-of-magnitude estimate based on differential rotation. Using the
solenoidal constraint and ignoring the radiation term, the induction equation reduces to
\(
\partial B_\phi/\partial t \simeq -(3/2)(v_\phi/r) B_r,
\)
corresponding to the standard $\Omega$--effect. For a $10\,M_\odot$ black hole at
$r\simeq3r_g$, the Keplerian shear rate is $v_\phi/r \sim 10^3\,\mathrm{s^{-1}}$.
Radiation-driven magentic fields of magnitude $B_r \sim 10^3$--$10^4\,\mathrm{G}$
therefore implies a toroidal-field growth rate
$\partial B_\phi/\partial t \sim 10^6$--$10^7\,\mathrm{G\,s^{-1}}$, leading to
$B_\phi \sim 10^6$--$10^7\,\mathrm{G}$ on a timescale of order one second. This
estimate is consistent with the peak toroidal field strengths observed in
Figures~\ref{lab_Gen_res} and \ref{lab_Gen_res_disk}.

In contrast, the radial and vertical
components remain largely radiation–dominated and show no comparable
amplification. 
Their evolution reflects the spatial structure of
the radiation source term rather than rotational shear.

The Hall term, when included, redistributes magnetic flux and couples field
components, but it does not by itself set the dominant field strength. Instead,
radiation acts primarily as a triggering mechanism, while azimuthal velocity provides the
dominant amplification channel. This selective enhancement of $B_\phi$ naturally
leads to the strong toroidal dominance observed and
demonstrates that even sub–Keplerian azimuthal motion can qualitatively alter the
magnetic–field morphology and strength in radiation–driven corona.
\section{Conclusions}
\label{sec_conc}
In this work, we have investigated the generation and amplification of magnetic
fields in the inner disc–corona environment of accreting black holes driven by
radiation forces and subsequent magnetohydrodynamic evolution. By explicitly
computing the radiation field for a realistic disc–corona geometry and coupling
its curl to Equation \ref{eq:hall_induction}, we have demonstrated that radiation
alone can produce dynamically relevant magnetic fields, which are then efficiently
amplified by rotational shear.

Our key result is that radiation–induced poloidal magnetic fields, although
initially modest, provide a strong field that is rapidly converted into a robust
toroidal component by differential rotation. For Eddington–level coronal
luminosities, the resulting toroidal magnetic field reaches strengths of
$B_\phi \sim 10^7 \,$G on timescales of order one second for a
$10\,M_\odot$ black hole. These values exceed simple equipartition estimates with the energy density. For an Eddington-level accretion flow at radii of a few $r_g$, the characteristic thermal energy density in the dense disc atmosphere corresponds to magnetic equipartition fields of order $B_{\rm eq,th} \sim 10^{5}$--$10^{6}\,\mathrm{G}$, as estimated in our previous work. The toroidal magnetic fields produced, reaching $B_\phi \sim 10^{7}\,\mathrm{G}$, therefore exceed thermal equipartition with the local gas pressure, reflecting the radiation-dominated and magnetically driven nature of the inner disc--corona environment. Unlike classical Poynting–Robertson batteries, which require many viscous times, the present mechanism operates on the local relevant viscous and advective timescale. 

A predominantly toroidal field can provide substantial magnetic pressure, enabling confinement and collimation of outflows and supplying the magnetic tension required for launching and stabilizing jets. Even in the absence of a large-scale pre-existing magnetic
field, the mechanism identified here naturally produces magnetic stresses
comparable to or exceeding the local radiation pressure and thermal pressure in
the inner corona. This opens a new pathway for magnetically driven winds and jet
formation in luminous accreting systems.

\subsection{Observational Implications}
The generation of strong magnetic fields triggered by radiation forces and subsequently amplified by differential rotation has direct and testable observational implications. In stellar-mass black hole systems, the amplified toroidal fields reach strengths of order $\sim 10^{7}$--$10^{8}$ G on dynamical timescales. Such magnetic fields can influence the geometry of the X-ray emitting region, the magnetization of disc-driven outflows, and the polarization properties of the observed radiation \citep{2020ARA&A..58..407D}.

\subsubsection*{X-ray Polarization and Faraday Effects}

A primary diagnostic of magnetic fields in the inner accretion flow and corona is X-ray polarimetry in the $\sim$2--8 keV band, as enabled by the Imaging X-ray Polarimetry Explorer (IXPE; \citealt{2022JATIS...8b6002W}). As polarized radiation propagates through magnetized plasma, it undergoes Faraday rotation, with the polarization angle scaling with the line-of-sight magnetic field component, the electron density, and the square of the wavelength.

Classical radiative–transfer calculations for magnetized, optically thick accretion–disc atmospheres show that once the Faraday rotation angle accumulated across an electron scattering optical depth of order unity becomes $\gtrsim 1$, the emergent polarization is strongly suppressed and originates only from a thin surface layer. For soft X-ray wavelengths, this condition corresponds to coherent magnetic fields of order $10^{6}$–$10^{7}\,\mathrm{G}$ within the scattering photosphere \citep{2002A&A...383..326S}.

Importantly, these constraints apply specifically to magnetic fields co-spatial with the X-ray emitting plasma and do not constrain the existence of stronger magnetic fields in regions immediately outside the photosphere or in vertically extended coronal and outflowing structures.

Further, magnetic fields produced in this work can be distinguished from MRI-driven fields by their large-scale coherent structure, in contrast to the stochastic, small-scale magnetic turbulence characteristic of MRI saturation \citep{2007A&A...476.1113F}.
Polarimetrically, such coherence would manifest as stable polarization angles and systematic, energy-dependent Faraday rotation, rather than the strong depolarization and rapidly varying polarization signatures expected from fully turbulent MRI-driven fields.
\subsubsection*{Jet Launching and Accretion States}

The emergence of strong toroidal magnetic fields naturally supplies the magnetic pressure and hoop stresses required for the launching and collimation of disc-driven winds and relativistic jets. While the present work does not model jet formation self-consistently, the field strengths obtained here are comparable to those typically invoked in magnetically arrested or magnetically dominated inner accretion flows \citep{1977MNRAS.179..433B, 2003PASJ...55L..69N}.

Spectral modelling of black hole accretion flows further suggests that magnetic pressure can approach or exceed gas pressure in hot, radiatively inefficient states, particularly in the inner regions of the flow \citep{2012ApJ...761..130Y}. Within this context, radiation-driven magnetic field generation offers a physically motivated mechanism for supplying the required magnetic flux locally, without the need for large-scale externally advected fields.

Overall, X-ray polarimetric observations provide critical empirical constraints that any magnetic-field generation mechanism must satisfy. The results presented here are consistent with these constraints if the strongest radiation-amplified magnetic fields reside predominantly outside the primary X-ray emitting photosphere, while still playing a dominant dynamical role in the surrounding corona and outflow regions.

The physical mechanism explored here is not restricted to steady accretion flows and can be naturally extended to high–energy transient sources, such as gamma–ray bursts, where intense, highly anisotropic radiation fields interact with rapidly evolving plasma outflows. Future work will investigate a fully relativistic generalization of this mechanism, incorporating special–relativistic plasma flows and the effects of curved spacetime, in order to assess radiation–triggered magnetic field generation in strongly relativistic environments.

\appendix
\section{Disc--corona geometry and emitting surfaces}
\label{sec_app}
The emitting configuration consists of two physically distinct components: a Keplerian
accretion disc and an extended coronal surface.
The disc photosphere is modelled as a horizontal surface located at $z=z_l=0$.
extending radially from the disc--corona interface at $x=x_c$ to an outer radius
$x_0 = 40\,r_g$.
The corona is represented by a single inclined emitting surface rather than by
separate top and side surfaces.
In the meridional $(r,z)$ plane, this surface is a straight line connecting the
points $(r=0,\,z=z_m)$ and $(r=x_c,\,z=z_l)$, where $z_m$ denotes the
maximum coronal height. We keep $z_l =0$.
Rotating this line about the symmetry axis generates an axisymmetric conical surface.
This geometry provides a smooth transition between the vertical and horizontal
extents of the corona and avoids the sharp edges inherent to a cylindrical model.

The coronal surface may be parametrized by a dimensionless coordinate
$s\in[0,1]$ as
\begin{equation}
r_s(s) = x_c\,s,
\qquad
z_s(s) = z_m - (z_m - z_l)\,s.
\end{equation}

The interior of the corona is assumed to be fully opaque.
Photons entering the region bounded by the disc photosphere and the coronal surface
are absorbed and do not contribute to the radiation field.
This assumption naturally enforces shadowing without the need for additional
visibility prescriptions.

\subsection{Photon propagation geometry}

The radiation field is evaluated at an arbitrary field point $(r,z)$ in cylindrical
coordinates.
A generic emitting surface element on the coronal surface is located at
$(r_s,\phi,z_s)$.
The separation between the surface element and the field point has magnitude
\begin{equation}
R =
\sqrt{
r^2 + r_s^2 + (z-z_s)^2 - 2 r r_s \cos\phi
}.
\end{equation}

The photon propagation direction $\hat{\boldsymbol{l}}$ has cylindrical components
\begin{equation}
l_r = \frac{r - r_s\cos\phi}{R},
\qquad
l_\phi = -\frac{r_s\sin\phi}{R},
\qquad
l_z = \frac{z - z_s}{R}.
\end{equation}

These expressions apply both to disc emission (with $z_s=z_l$ and $r_s=x$) and to
coronal emission (with $r_s=r_s(s)$ and $z_s=z_s(s)$).

\subsection{Surface projection and visibility}

Emission from all surfaces is assumed to be locally Lambertian in comoving frame, where the emission is isotropic from the surface.
However, each surface element contributes a projection factor
$(\hat{\boldsymbol{n}}\cdot\hat{\boldsymbol{l}})$, where
$\hat{\boldsymbol{n}}$ is the outward surface normal. The inward emitted photons are thus excluded.

For the disc photosphere, which is horizontal, the normal vector is $\hat{z}$,
and the projection factor reduces to $l_z$.
For the coronal surface, the outward unit normal lies in the $(r,z)$ plane and is
constant along the surface,
\begin{equation}
\hat{\boldsymbol{n}}_{\rm c}
=
\frac{1}{\sqrt{x_c^2 + (z_m - z_l)^2}}
\left[
(z_m - z_l)\,\hat{r}
-
x_c\,\hat{z}
\right].
\end{equation}
The corresponding projection factor is
\begin{equation}
\hat{\boldsymbol{n}}_{\rm c}\cdot\hat{\boldsymbol{l}}
=
n_r\,l_r + n_z\,l_z.
\end{equation}

Emission from the coronal surface is included only when this quantity is positive,
ensuring that photons are emitted outward.
Combined with the opacity of the coronal interior, this condition guarantees that
radiation reaches the surrounding plasma only along geometrically allowed paths.

\subsection{Relativistic Doppler effects}

All emitting surfaces are assumed to rotate azimuthally.
The disc photosphere rotates with the local Keplerian angular velocity,
\begin{equation}
\Omega_{\rm K}(x)
=
\left[
\frac{x}{2(x-1)^2}
\right]^{1/2}.
\end{equation}

The corona is assumed to rotate differentially along its surface, with the local
azimuthal speed determined by the cylindrical radius $r_s$,
\begin{equation}
v_\phi(r_s)
=
\begin{cases}
\Omega_{\rm K}(x_c)\,r_s/x_c, & r_s < x_c, \\[6pt]
\Omega_{\rm K}(r_s)\,r_s, & r_s \ge x_c .
\end{cases}
\end{equation}
This prescription ensures continuity of the azimuthal velocity at the disc--corona
interface.

Special relativistic effects are included through Doppler boosting.
The Doppler factor enters the radiation field as
\begin{equation}
\mathcal{D}^{-4}
=
\gamma^2
\left(
1 - \boldsymbol{v}\cdot\hat{\boldsymbol{l}}
\right)^4,
\qquad
\gamma = (1 - v_\phi^2)^{-1/2},
\end{equation}
with $\boldsymbol{v} = v_\phi\,\hat{\phi}$.

\subsection{Radiation flux integrals}

The radiation flux components $(F_r,F_\phi,F_z)$ at the field point $(r,z)$ are
obtained by integrating over all emitting surfaces,
\begin{equation}
F_i(r,z)
=
\int
\frac{
I(s)\,
(\hat{\boldsymbol{n}}\cdot\hat{\boldsymbol{l}})\,
l_i
}{
R^3\,\mathcal{D}^4
}
\,\mathrm{d}A,
\end{equation}
where $I$ is the surface emissivity.

The radiation field is constructed from the superposition of emission originating from the Keplerian disc photosphere and from the inclined coronal surface described above. The specific intensity of the Keplerian disc is assumed to follow the standard Shakura--Sunyaev prescription,
\begin{equation}
I_{\rm d}(x)
=
\frac{3GM\dot{M}}{8\pi x^3}
\left(
1 - \sqrt{\frac{x_{\rm in}}{x}}
\right),
\label{eq:Idisk}
\end{equation}
where $x$ is the cylindrical radius measured in physical units, $x_{\rm in}$ is the inner truncation radius of the Keplerian disc, and $\dot{M}$ is the mass accretion rate. We parametrize the accretion rate as
\begin{equation}
\dot{M} = \dot{m}\,\dot{M}_{\rm Edd},
\end{equation}
with $\dot{M}_{\rm Edd}$ being the Eddington accretion rate and $\dot{m}$ the dimensionless accretion rate. Throughout this work, the disc is truncated at $x_{\rm in}\simeq 3\,r_g$, consistent with the innermost stable circular orbit for a non-rotating black hole.

In contrast to the Keplerian disc, the coronal emission is assumed to be homogeneous over its emitting surface. This assumption is motivated by global radiation-MHD simulations, which show that compact inner coronae are characterized by relatively uniform radiation energy density and flux distributions. We therefore prescribe a constant specific intensity over the entire coronal surface,
\begin{equation}
I_{\rm c}
=
\frac{L_{\rm c}}{A_{\rm c}\,c},
\label{eq:Icorona}
\end{equation}
where $L_{\rm c}$ is the total coronal luminosity and $A_{\rm c}$ is the total emitting area of the coronal surface.

For the disc, the surface element is
\begin{equation}
\mathrm{d}A = x\,\mathrm{d}x\,\mathrm{d}\phi.
\end{equation}
The coronal emitting surface is parametrized by a dimensionless coordinate $s \in [0,1]$, which labels points along the inclined coronal surface. The surface is defined in the meridional $(r,z)$ plane by the parametric functions $r_s(s)$ and $z_s(s)$, such that $s=0$ corresponds to the apex of the corona at $(r=0, z=z_m)$ and $s=1$ corresponds to the disc--corona interface at $(r=x_c, z=z_l)$. Rotating this curve about the symmetry axis generates the axisymmetric coronal emitting surface.
For the coronal surface, the surface element is
\begin{equation}
\mathrm{d}A
=
r_s(s)\,
\sqrt{
\left(\frac{\mathrm{d}r_s}{\mathrm{d}s}\right)^2
+
\left(\frac{\mathrm{d}z_s}{\mathrm{d}s}\right)^2
}
\,\mathrm{d}s\,\mathrm{d}\phi.
\end{equation}

The total radiation flux is obtained by summing the disc and coronal contributions.
All flux components are set identically to zero inside the optically thick disc and
coronal regions. If corona is considered to be optically thin, the radiative output would require volume emissivity and internal transport, requiring detailed radiative transport compared to surface-based emissivity considered here.

\subsection{Construction of the radiation--driven source term}

The radiation driven magnetic source term entering the induction equation is defined
as
\begin{equation}
\mathbf{F}_0(r,z)
=
-\frac{\sigma_T}{e}\,
\nabla\times\mathbf{F}_{\rm rad}(r,z),
\end{equation}
where $\mathbf{F}_{\rm rad}$ is the total radiation flux computed as described above.

Spatial derivatives of the flux components are evaluated numerically on a
high--resolution $(r,z)$ grid.
The resulting source term $\mathbf{F}_0(r,z)$ is stored as a two--dimensional table
and interpolated onto the computation domain during the magnetic-field evolution.

By construction, $\mathbf{F}_0$ vanishes identically inside the optically thick disc
and corona, and is non-zero only in the surrounding plasma.
This allows the subsequent magnetic induction and Hall evolution to be studied in
isolation from radiative transfer effects. This construction reproduces the radiation field used in \cite{2025ApJ...988L..59V} with vertically extended corona.
For an elaborated description of radiation field evaluation around black holes, see \citep{2008bhad.book.....K, 2005MNRAS.356..145C}.

\section*{Acknowledgments}
We acknowledge the support from the European Union (EU) via ERC consolidator grant 773062 (O.M.J.) and from the Israel Space Agency via grant \#6766. 

\bibliography{example}{}

@ARTICLE{2010ApJ...716.1566A,
       author = {{Ando}, Masashi and {Doi}, Kentaro and {Susa}, Hajime},
        title = "{Generation of Seed Magnetic Field Around First Stars: Effects of Radiation Force}",
      journal = {\apj},
         year = 2010,
        month = jun,
       volume = {716},
       number = {2},
        pages = {1566-1572},
          doi = {10.1088/0004-637X/716/2/1566},
archivePrefix = {arXiv},
       eprint = {1005.1123},
 primaryClass = {astro-ph.CO},
       adsurl = {https://ui.adsabs.harvard.edu/abs/2010ApJ...716.1566A}
}

@ARTICLE{2014ApJ...782..108S,
       author = {{Shiromoto}, Yuki and {Susa}, Hajime and {Hosokawa}, Takashi},
        title = "{Generation of Magnetic Field on the Accretion Disk around a Proto-first-star}",
      journal = {\apj},
         year = 2014,
        month = feb,
       volume = {782},
       number = {2},
          eid = {108},
        pages = {108},
          doi = {10.1088/0004-637X/782/2/108},
archivePrefix = {arXiv},
       eprint = {1401.0905},
 primaryClass = {astro-ph.CO},
       adsurl = {https://ui.adsabs.harvard.edu/abs/2014ApJ...782..108S}
}

@ARTICLE{1977A&A....59..111B,
       author = {{Bisnovatyi-Kogan}, G.~S. and {Blinnikov}, S.~I.},
        title = "{Disk accretion onto a black hole at subcritical luminosity.}",
      journal = {\aap},
         year = 1977,
        month = jul,
       volume = {59},
        pages = {111-125},
       adsurl = {https://ui.adsabs.harvard.edu/abs/1977A&A....59..111B}
}

@ARTICLE{1998ApJ...508..859C,
       author = {{Contopoulos}, Ioannis and {Kazanas}, Demosthenes},
        title = "{A Cosmic Battery}",
      journal = {\apj},
         year = 1998,
        month = dec,
       volume = {508},
       number = {2},
        pages = {859-863},
          doi = {10.1086/306426},
archivePrefix = {arXiv},
       eprint = {astro-ph/9808223},
 primaryClass = {astro-ph},
       adsurl = {https://ui.adsabs.harvard.edu/abs/1998ApJ...508..859C}
}

@ARTICLE{2002ApJ...580..380B,
       author = {{Bisnovatyi-Kogan}, G.~S. and {Lovelace}, R.~V.~E. and {Belinski}, V.~A.},
        title = "{A Cosmic Battery Reconsidered}",
      journal = {\apj},
         year = 2002,
        month = nov,
       volume = {580},
       number = {1},
        pages = {380-388},
          doi = {10.1086/342876},
archivePrefix = {arXiv},
       eprint = {astro-ph/0207476},
 primaryClass = {astro-ph},
       adsurl = {https://ui.adsabs.harvard.edu/abs/2002ApJ...580..380B}
}

@ARTICLE{2006ApJ...652.1451C,
       author = {{Contopoulos}, Ioannis and {Kazanas}, Demosthenes and {Christodoulou}, Dimitris M.},
        title = "{The Cosmic Battery Revisited}",
      journal = {\apj},
         year = 2006,
        month = dec,
       volume = {652},
       number = {2},
        pages = {1451-1456},
          doi = {10.1086/507600},
archivePrefix = {arXiv},
       eprint = {astro-ph/0608701},
 primaryClass = {astro-ph},
       adsurl = {https://ui.adsabs.harvard.edu/abs/2006ApJ...652.1451C}
}

@ARTICLE{2008ApJ...674..388C,
       author = {{Christodoulou}, Dimitris M. and {Contopoulos}, Ioannis and {Kazanas}, Demosthenes},
        title = "{Simulations of the Poynting-Robertson Cosmic Battery in Resistive Accretion Disks}",
      journal = {\apj},
         year = 2008,
        month = feb,
       volume = {674},
       number = {1},
        pages = {388-407},
          doi = {10.1086/524699},
archivePrefix = {arXiv},
       eprint = {0706.3187},
 primaryClass = {astro-ph},
       adsurl = {https://ui.adsabs.harvard.edu/abs/2008ApJ...674..388C}
}

@ARTICLE{2003PhRvD..67d3505L,
       author = {{Langer}, Mathieu and {Puget}, Jean-Loup and {Aghanim}, Nabila},
        title = "{Cosmological magnetogenesis driven by radiation pressure}",
      journal = {\prd},
         year = 2003,
        month = feb,
       volume = {67},
       number = {4},
          eid = {043505},
        pages = {043505},
          doi = {10.1103/PhysRevD.67.043505},
archivePrefix = {arXiv},
       eprint = {astro-ph/0212108},
 primaryClass = {astro-ph},
       adsurl = {https://ui.adsabs.harvard.edu/abs/2003PhRvD..67d3505L}
}

@ARTICLE{2021MNRAS.501.4850R,
       author = {{Raychaudhuri}, Sananda and {Vyas}, Mukesh K. and {Chattopadhyay}, Indranil},
        title = "{Simulations of radiation-driven winds from Keplerian discs}",
      journal = {\mnras},
         year = 2021,
        month = mar,
       volume = {501},
       number = {4},
        pages = {4850-4860},
          doi = {10.1093/mnras/staa3920},
archivePrefix = {arXiv},
       eprint = {2012.08886},
 primaryClass = {astro-ph.HE},
       adsurl = {https://ui.adsabs.harvard.edu/abs/2021MNRAS.501.4850R}
}

@ARTICLE{2019MNRAS.482.4203V,
       author = {{Vyas}, Mukesh K. and {Chattopadhyay}, Indranil},
        title = "{Radiation driving and heating of general relativistic jets under a Compton-scattering regime}",
      journal = {MNRAS},
         year = 2019,
        month = jan,
       volume = {482},
       number = {3},
        pages = {4203-4214},
          doi = {10.1093/mnras/sty2917},
archivePrefix = {arXiv},
       eprint = {1810.11183},
 primaryClass = {astro-ph.HE},
       adsurl = {https://ui.adsabs.harvard.edu/abs/2019MNRAS.482.4203V}
}

@BOOK{1988ASSL..133.....R,
       author = {{Ruzmaikin}, Aleksandr A. and {Sokolov}, Dmitrii D. and {Shukurov}, Anvar M.},
        title = "{Magnetic Fields of Galaxies}",
         year = 1988,
       volume = {133},
          doi = {10.1007/978-94-009-2835-0},
       adsurl = {https://ui.adsabs.harvard.edu/abs/1988ASSL..133.....R}
}

@ARTICLE{2005PPCF...47A.205S,
       author = {{Schlickeiser}, Reinhard},
        title = "{On the origin of cosmological magnetic fields by plasma instabilities}",
      journal = {Plasma Physics and Controlled Fusion},
         year = 2005,
        month = may,
       volume = {47},
       number = {5A},
        pages = {A205-A218},
          doi = {10.1088/0741-3335/47/5A/015},
       adsurl = {https://ui.adsabs.harvard.edu/abs/2005PPCF...47A.205S}
}

@ARTICLE{2002RvMP...74..775W,
       author = {{Widrow}, Lawrence M.},
        title = "{Origin of galactic and extragalactic magnetic fields}",
      journal = {Reviews of Modern Physics},
         year = 2002,
        month = jan,
       volume = {74},
       number = {3},
        pages = {775-823},
          doi = {10.1103/RevModPhys.74.775},
archivePrefix = {arXiv},
       eprint = {astro-ph/0207240},
 primaryClass = {astro-ph},
       adsurl = {https://ui.adsabs.harvard.edu/abs/2002RvMP...74..775W}
}

@ARTICLE{2002PhT....55l..40K,
       author = {{Kronberg}, Philipp P.},
        title = "{Intergalactic magnetic fields.}",
      journal = {Physics Today},
         year = 2002,
        month = dec,
       volume = {55},
       number = {12},
        pages = {12.40},
          doi = {10.1063/1.1537911},
       adsurl = {https://ui.adsabs.harvard.edu/abs/2002PhT....55l..40K}
}

@BOOK{2008bhad.book.....K,
       author = {{Kato}, S. and {Fukue}, J. and {Mineshige}, S.},
        title = "{Black-Hole Accretion Disks --- Towards a New Paradigm ---}",
         year = 2008,
       adsurl = {https://ui.adsabs.harvard.edu/abs/2008bhad.book.....K}
}

@ARTICLE{2023ApJS..264...32B,
       author = {{B{\'e}gu{\'e}}, D. and {Pe'er}, A. and {Zhang}, G. -Q. and {Zhang}, B. -B. and {Pevzner}, B.},
        title = "{cuHARM: A New GPU-accelerated GRMHD Code and Its Application to ADAF Disks}",
      journal = {\apjs},
         year = 2023,
        month = feb,
       volume = {264},
       number = {2},
          eid = {32},
        pages = {32},
          doi = {10.3847/1538-4365/aca276},
archivePrefix = {arXiv},
       eprint = {2205.02484},
 primaryClass = {astro-ph.HE},
       adsurl = {https://ui.adsabs.harvard.edu/abs/2023ApJS..264...32B}
}

@ARTICLE{2020ARA&A..58..407D, 
       author = {{Davis}, Shane W. and {Tchekhovskoy}, Alexander},
        title = "{Magnetohydrodynamics Simulations of Active Galactic Nucleus Disks and Jets}",
      journal = {\araa},
         year = 2020,
        month = aug,
       volume = {58},
        pages = {407-439},
          doi = {10.1146/annurev-astro-081817-051905},
archivePrefix = {arXiv},
       eprint = {2101.08839},
 primaryClass = {astro-ph.HE},
       adsurl = {https://ui.adsabs.harvard.edu/abs/2020ARA&A..58..407D}
}

@ARTICLE{2021NewAR..9201610K,
       author = {{Komissarov}, Serguei and {Porth}, Oliver},
        title = "{Numerical simulations of jets}",
      journal = {\nar},
         year = 2021,
        month = jun,
       volume = {92},
          eid = {101610},
        pages = {101610},
          doi = {10.1016/j.newar.2021.101610},
       adsurl = {https://ui.adsabs.harvard.edu/abs/2021NewAR..9201610K}
}

@ARTICLE{2022MNRAS.511.3795N,
       author = {{Narayan}, Ramesh and {Chael}, Andrew and {Chatterjee}, Koushik and {Ricarte}, Angelo and {Curd}, Brandon},
        title = "{Jets in magnetically arrested hot accretion flows: geometry, power, and black hole spin-down}",
      journal = {\mnras},
         year = 2022,
        month = apr,
       volume = {511},
       number = {3},
        pages = {3795-3813},
          doi = {10.1093/mnras/stac285},
archivePrefix = {arXiv},
       eprint = {2108.12380},
 primaryClass = {astro-ph.HE},
       adsurl = {https://ui.adsabs.harvard.edu/abs/2022MNRAS.511.3795N}
}

@article{Miniutti2004,
  author = {Miniutti, G. and Fabian, A. C.},
  title = {A light bending model for the X-ray temporal and spectral properties of accreting black holes},
  journal = {Monthly Notices of the Royal Astronomical Society},
  volume = {349},
  issue = {4},
  pages = {1435-1448},
  year = {2004},
  doi = {10.1111/j.1365-2966.2004.07611.x},
  url = {https://academic.oup.com/mnras/article/349/4/1435/1012057}
}

@ARTICLE{1991ApJ...376..214B,
       author = {{Balbus}, Steven A. and {Hawley}, John F.},
        title = "{A Powerful Local Shear Instability in Weakly Magnetized Disks. I. Linear Analysis}",
      journal = {\apj},
         year = 1991,
        month = jul,
       volume = {376},
        pages = {214},
          doi = {10.1086/170270},
       adsurl = {https://ui.adsabs.harvard.edu/abs/1991ApJ...376..214B}
}

@ARTICLE{1995ApJ...440..742H,
       author = {{Hawley}, John F. and {Gammie}, Charles F. and {Balbus}, Steven A.},
        title = "{Local Three-dimensional Magnetohydrodynamic Simulations of Accretion Disks}",
      journal = {\apj},
         year = 1995,
        month = feb,
       volume = {440},
        pages = {742},
          doi = {10.1086/175311},
       adsurl = {https://ui.adsabs.harvard.edu/abs/1995ApJ...440..742H}
}

@ARTICLE{2005MNRAS.356..145C,
       author = {{Chattopadhyay}, Indranil},
        title = "{Radiatively driven rotating pair-plasma jets from two-component accretion flows}",
      journal = {\mnras},
         year = 2005,
        month = jan,
       volume = {356},
       number = {1},
        pages = {145-166},
          doi = {10.1111/j.1365-2966.2004.08429.x},
archivePrefix = {arXiv},
       eprint = {astro-ph/0411224},
 primaryClass = {astro-ph},
       adsurl = {https://ui.adsabs.harvard.edu/abs/2005MNRAS.356..145C}
}

@BOOK{1978mfge.book.....M,
       author = {{Moffatt}, H.~K.},
        title = "{Magnetic field generation in electrically conducting fluids}",
         year = 1978,
       adsurl = {https://ui.adsabs.harvard.edu/abs/1978mfge.book.....M}
}

@ARTICLE{2015ApJ...805..105C,
       author = {{Contopoulos}, Ioannis and {Nathanail}, Antonios and {Katsanikas}, Matthaios},
        title = "{The Cosmic Battery in Astrophysical Accretion Disks}",
      journal = {\apj},
         year = 2015,
        month = jun,
       volume = {805},
       number = {2},
          eid = {105},
        pages = {105},
          doi = {10.1088/0004-637X/805/2/105},
archivePrefix = {arXiv},
       eprint = {1501.05784},
 primaryClass = {astro-ph.HE},
       adsurl = {https://ui.adsabs.harvard.edu/abs/2015ApJ...805..105C}
}

@ARTICLE{2014ApJ...794...27K,
       author = {{Koutsantoniou}, Leela E. and {Contopoulos}, Ioannis},
        title = "{Accretion Disk Radiation Dynamics and the Cosmic Battery}",
      journal = {\apj},
         year = 2014,
        month = oct,
       volume = {794},
       number = {1},
          eid = {27},
        pages = {27},
          doi = {10.1088/0004-637X/794/1/27},
archivePrefix = {arXiv},
       eprint = {1405.6018},
 primaryClass = {astro-ph.HE},
       adsurl = {https://ui.adsabs.harvard.edu/abs/2014ApJ...794...27K}
}

@ARTICLE{2012A&A...537A..18M,
       author = {{Mineo}, T. and {Massaro}, E. and {D'Ai}, A. and {Massa}, F. and {Feroci}, M. and {Ventura}, G. and {Casella}, P. and {Ferrigno}, C. and {Belloni}, T.},
        title = "{The complex behaviour of the microquasar GRS 1915+105 in the {\ensuremath{\rho}} class observed with BeppoSAX. II. Time-resolved spectral analysis}",
      journal = {\aap},
         year = 2012,
        month = jan,
       volume = {537},
          eid = {A18},
        pages = {A18},
          doi = {10.1051/0004-6361/201117369},
archivePrefix = {arXiv},
       eprint = {1110.5199},
 primaryClass = {astro-ph.SR},
       adsurl = {https://ui.adsabs.harvard.edu/abs/2012A&A...537A..18M}
}

@ARTICLE{2025ApJ...988L..59V,
       author = {{Vyas}, Mukesh Kumar and {Pe'er}, Asaf},
        title = "{Generation of Magnetic Fields around Black Hole Accretion Disks due to Nonconservative Radiation Fields}",
      journal = {\apjl},
         year = 2025,
        month = aug,
       volume = {988},
       number = {2},
          eid = {L59},
        pages = {L59},
          doi = {10.3847/2041-8213/aded11},
archivePrefix = {arXiv},
       eprint = {2505.10460},
 primaryClass = {astro-ph.HE},
       adsurl = {https://ui.adsabs.harvard.edu/abs/2025ApJ...988L..59V}
}

@ARTICLE{1958ApJ...128..664P,
       author = {{Parker}, E.~N.},
        title = "{Dynamics of the Interplanetary Gas and Magnetic Fields.}",
      journal = {\apj},
         year = 1958,
        month = nov,
       volume = {128},
        pages = {664},
          doi = {10.1086/146579},
       adsurl = {https://ui.adsabs.harvard.edu/abs/1958ApJ...128..664P}
}

@ARTICLE{2002A&A...383..326S,
       author = {{Silant'ev}, N.~A.},
        title = "{Polarization from magnetized optically thick accretion disks}",
      journal = {\aap},
         year = 2002,
        month = jan,
       volume = {383},
        pages = {326-337},
          doi = {10.1051/0004-6361:20011740},
       adsurl = {https://ui.adsabs.harvard.edu/abs/2002A&A...383..326S}
}

@ARTICLE{2022JATIS...8b6002W,
       author = {{Weisskopf}, Martin C. and {Soffitta}, Paolo and {Baldini}, Luca and {Ramsey}, Brian D. and {O'Dell}, Stephen L. and {Romani}, Roger W. and {Matt}, Giorgio and {Deininger}, William D. and {Baumgartner}, Wayne H. and {Bellazzini}, Ronaldo and {Costa}, Enrico and {Kolodziejczak}, Jeffery J. and {Latronico}, Luca and {Marshall}, Herman L. and {Muleri}, Fabio and {Bongiorno}, Stephen D. and {Tennant}, Allyn and {Bucciantini}, Niccolo and {Dovciak}, Michal and {Marin}, Frederic and {Marscher}, Alan and {Poutanen}, Juri and {Slane}, Pat and {Turolla}, Roberto and {Kalinowski}, William and {Di Marco}, Alessandro and {Fabiani}, Sergio and {Minuti}, Massimo and {La Monaca}, Fabio and {Pinchera}, Michele and {Rankin}, John and {Sgro'}, Carmelo and {Trois}, Alessio and {Xie}, Fei and {Alexander}, Cheryl and {Allen}, D. Zachery and {Amici}, Fabrizio and {Andersen}, Jason and {Antonelli}, Angelo and {Antoniak}, Spencer and {Attin{\`a}}, Primo and {Barbanera}, Mattia and {Bachetti}, Matteo and {Baggett}, Randy M. and {Bladt}, Jeff and {Brez}, Alessandro and {Bonino}, Raffaella and {Boree}, Christopher and {Borotto}, Fabio and {Breeding}, Shawn and {Brienza}, Daniele and {Bygott}, H. Kyle and {Caporale}, Ciro and {Cardelli}, Claudia and {Carpentiero}, Rita and {Castellano}, Simone and {Castronuovo}, Marco and {Cavalli}, Luca and {Cavazzuti}, Elisabetta and {Ceccanti}, Marco and {Centrone}, Mauro and {Citraro}, Saverio and {D'Amico}, Fabio and {D'Alba}, Elisa and {Di Gesu}, Laura and {Del Monte}, Ettore and {Dietz}, Kurtis L. and {Di Lalla}, Niccolo' and {Persio}, Giuseppe Di and {Dolan}, David and {Donnarumma}, Immacolata and {Evangelista}, Yuri and {Ferrant}, Kevin and {Ferrazzoli}, Riccardo and {Ferrie}, MacKenzie and {Footdale}, Joseph and {Forsyth}, Brent and {Foster}, Michelle and {Garelick}, Benjamin and {Gunji}, Shuichi and {Gurnee}, Eli and {Head}, Michael and {Hibbard}, Grant and {Johnson}, Samantha and {Kelly}, Erik and {Kilaru}, Kiranmayee and {Lefevre}, Carlo and {Roy}, Shelley Le and {Loffredo}, Pasqualino and {Lorenzi}, Paolo and {Lucchesi}, Leonardo and {Maddox}, Tyler and {Magazzu}, Guido and {Maldera}, Simone and {Manfreda}, Alberto and {Mangraviti}, Elio and {Marengo}, Marco and {Marrocchesi}, Alessandra and {Massaro}, Francesco and {Mauger}, David and {McCracken}, Jeffrey and {McEachen}, Michael and {Mize}, Rondal and {Mereu}, Paolo and {Mitchell}, Scott and {Mitsuishi}, Ikuyuki and {Morbidini}, Alfredo and {Mosti}, Federico and {Nasimi}, Hikmat and {Negri}, Barbara and {Negro}, Michela and {Nguyen}, Toan and {Nitschke}, Isaac and {Nuti}, Alessio and {Onizuka}, Mitch and {Oppedisano}, Chiara and {Orsini}, Leonardo and {Osborne}, Darren and {Pacheco}, Richard and {Paggi}, Alessandro and {Painter}, Will and {Pavelitz}, Steven D. and {Pentz}, Christina and {Piazzolla}, Raffaele and {Perri}, Matteo and {Pesce-Rollins}, Melissa and {Peterson}, Colin and {Pilia}, Maura and {Profeti}, Alessandro and {Puccetti}, Simonetta and {Ranganathan}, Jaganathan and {Ratheesh}, Ajay and {Reedy}, Lee and {Root}, Noah and {Rubini}, Alda and {Ruswick}, Stephanie and {Sanchez}, Javier and {Sarra}, Paolo and {Santoli}, Francesco and {Scalise}, Emanuele and {Sciortino}, Andrea and {Schroeder}, Christopher and {Seek}, Tim and {Sosdian}, Kalie and {Spandre}, Gloria and {Speegle}, Chet O. and {Tamagawa}, Toru and {Tardiola}, Marcello and {Tobia}, Antonino and {Thomas}, Nicholas E. and {Valerie}, Robert and {Vimercati}, Marco and {Walden}, Amy L. and {Weddendorf}, Bruce and {Wedmore}, Jeffrey and {Welch}, David and {Zanetti}, Davide and {Zanetti}, Francesco},
        title = "{The Imaging X-Ray Polarimetry Explorer (IXPE): Pre-Launch}",
      journal = {Journal of Astronomical Telescopes, Instruments, and Systems},
         year = 2022,
        month = apr,
       volume = {8},
       number = {2},
          eid = {026002},
        pages = {026002},
          doi = {10.1117/1.JATIS.8.2.026002},
archivePrefix = {arXiv},
       eprint = {2112.01269},
 primaryClass = {astro-ph.IM},
       adsurl = {https://ui.adsabs.harvard.edu/abs/2022JATIS...8b6002W}
}

@ARTICLE{1977MNRAS.179..433B,
       author = {{Blandford}, R.~D. and {Znajek}, R.~L.},
        title = "{Electromagnetic extraction of energy from Kerr black holes.}",
      journal = {\mnras},
         year = 1977,
        month = may,
       volume = {179},
        pages = {433-456},
          doi = {10.1093/mnras/179.3.433},
       adsurl = {https://ui.adsabs.harvard.edu/abs/1977MNRAS.179..433B}
}

@ARTICLE{2003PASJ...55L..69N,
       author = {{Narayan}, Ramesh and {Igumenshchev}, Igor V. and {Abramowicz}, Marek A.},
        title = "{Magnetically Arrested Disk: an Energetically Efficient Accretion Flow}",
      journal = {\pasj},
         year = 2003,
        month = dec,
       volume = {55},
        pages = {L69-L72},
          doi = {10.1093/pasj/55.6.L69},
archivePrefix = {arXiv},
       eprint = {astro-ph/0305029},
 primaryClass = {astro-ph},
       adsurl = {https://ui.adsabs.harvard.edu/abs/2003PASJ...55L..69N}
}

@ARTICLE{2012ApJ...761..130Y,
       author = {{Yuan}, Feng and {Bu}, Defu and {Wu}, Maochun},
        title = "{Numerical Simulation of Hot Accretion Flows. II. Nature, Origin, and Properties of Outflows and their Possible Observational Applications}",
      journal = {\apj},
         year = 2012,
        month = dec,
       volume = {761},
       number = {2},
          eid = {130},
        pages = {130},
          doi = {10.1088/0004-637X/761/2/130},
archivePrefix = {arXiv},
       eprint = {1206.4173},
 primaryClass = {astro-ph.HE},
       adsurl = {https://ui.adsabs.harvard.edu/abs/2012ApJ...761..130Y}
}

@book{schiesser2012numerical,
  title={The numerical method of lines: integration of partial differential equations},
  author={Schiesser, William E},
  year={2012},
  publisher={Elsevier}
}

@ARTICLE{1976SvAL....2..191B,
       author = {{Bisnovatyi-Kogan}, G.~S. and {Blinnikov}, S.~I.},
        title = "{A hot corona around a black-hole accretion disk as a model for CYG X-1.}",
      journal = {Soviet Astronomy Letters},
         year = 1976,
        month = oct,
       volume = {2},
        pages = {191-193},
          doi = {10.48550/arXiv.astro-ph/0003275},
archivePrefix = {arXiv},
       eprint = {astro-ph/0003275},
 primaryClass = {astro-ph},
       adsurl = {https://ui.adsabs.harvard.edu/abs/1976SvAL....2..191B}
}

@ARTICLE{1991ApJ...380L..51H,
       author = {{Haardt}, F. and {Maraschi}, L.},
        title = "{A Two-Phase Model for the X-Ray Emission from Seyfert Galaxies}",
      journal = {\apjl},
         year = 1991,
        month = oct,
       volume = {380},
        pages = {L51},
          doi = {10.1086/186171},
       adsurl = {https://ui.adsabs.harvard.edu/abs/1991ApJ...380L..51H}
}

@ARTICLE{2004MNRAS.355.1105F,
       author = {{Fender}, R.~P. and {Belloni}, T.~M. and {Gallo}, E.},
        title = "{Towards a unified model for black hole X-ray binary jets}",
      journal = {\mnras},
         year = 2004,
        month = dec,
       volume = {355},
       number = {4},
        pages = {1105-1118},
          doi = {10.1111/j.1365-2966.2004.08384.x},
archivePrefix = {arXiv},
       eprint = {astro-ph/0409360},
 primaryClass = {astro-ph},
       adsurl = {https://ui.adsabs.harvard.edu/abs/2004MNRAS.355.1105F}
}

@ARTICLE{2007A&A...476.1113F,
       author = {{Fromang}, S. and {Papaloizou}, J.},
        title = "{MHD simulations of the magnetorotational instability in a shearing box with zero net flux. I. The issue of convergence}",
      journal = {\aap},
         year = 2007,
        month = dec,
       volume = {476},
       number = {3},
        pages = {1113-1122},
          doi = {10.1051/0004-6361:20077942},
archivePrefix = {arXiv},
       eprint = {0705.3621},
 primaryClass = {astro-ph},
       adsurl = {https://ui.adsabs.harvard.edu/abs/2007A&A...476.1113F}
}
\bibliographystyle{aasjournal}
\end{document}